\documentclass[a4paper]{jpconf}
\usepackage[utf8]{inputenc}
\usepackage{amsmath,amsfonts,amssymb,amscd,braket}
\usepackage{graphicx}
\usepackage{tikz}  
\usetikzlibrary{positioning,matrix,shapes,shapes.callouts,decorations.pathmorphing,shapes.misc,shapes.geometric,arrows,external,shadows,shapes.symbols}
\usepackage{cancel}
\usepackage{mathrsfs}
\usepackage{bbm}
\usepackage{xspace} 
\usepackage[small,loose,md,FIGTOPCAP]{subfigure}  
\usepackage{multicol}
\usepackage{multirow} 
\usepackage{euscript} 

\newcommand{\I}{\mathrm{i}}

\newcommand{\SO}[1]{\ensuremath{\mathrm{SO}(#1)}}
\newcommand{\SU}[1]{\ensuremath{\mathrm{SU}(#1)}}

\newcommand{\Z}[1]{\ensuremath{\mathbbm{Z}_{#1}}} 

\newcommand{\T}[1]{\ensuremath{\mathrm{T}_{#1}}}

\newcommand*{\rep}[2][]{\ensuremath{{\boldsymbol{#2}#1}}} 

\renewcommand{\bar}[1]{\overline{#1}}
\newcommand{\elm}[1]{\mathsf{#1}}

\newcommand{\parityP}{\ensuremath{\EuScript P}}
\newcommand{\Out}{\ensuremath{\mathrm{Out}}\xspace}
\newcommand{\Outs}{\ensuremath{\mathrm{Outs}}\xspace}

\newcommand{\Tabref}[1]{table~\ref{#1}}
\newcommand{\Secref}[1]{section~\ref{#1}}

\usepackage[bookmarks=false]{hyperref}

\tikzstyle{every picture}+=[remember picture] 
\everymath{\displaystyle} 

\begin{document}
\title{CP as a Symmetry of Symmetries}

\author{Andreas Trautner}

\address{Bethe Center for Theoretical Physics und Physikalisches Institut der Universit\"at Bonn, Nussallee 12, 53115 Bonn, Germany}

\ead{atrautner@uni-bonn.de}

\begin{abstract}
It is explained that the Standard Model combined charge conjugation and parity transformation (CP) 
is a simultaneous complex conjugation outer automorphism transformation of gauge and space-time symmetries.
Simple examples are given for the general concept of outer automorphisms (``symmetries of symmetries''), as well 
as for their possible actions on physical theories.
It is highlighted that complex conjugation outer automorphisms do not, in general, exist for all symmetries. Examples are given
for cases in which the physical CP transformation is violated as a consequence of requiring another symmetry.
A toy model is illustrated in which CP is spontaneously violated in the broken phase of a continuous gauge symmetry, 
while an unbroken outer automorphism protects the topological vacuum angle at $\theta=0$.
\end{abstract}

\section{Introduction}
One of the main questions raised by the Standard Model of particle physics (SM)
is the origin of the SM flavor structure.
By observation, the violation of flavor and the violation of matter-anti matter symmetry are highly intertwined in the SM.
In particular, CP violation has never been observed without flavor violation, 
implying that also the strong CP problem can be regarded as inherent part of the flavor puzzle. 
Ultimately, a complete theory of flavor should also be a theory of CP violation.

Among many other approaches, the flavor puzzle has been addressed with discrete symmetries. 
In this context, the interplay of CP transformations and discrete flavor symmetries has been investigated \cite{Feruglio:2012cw,Holthausen:2012dk}.
It has been found that CP transformations are described by particular outer automorphisms of discrete symmetries \cite{Holthausen:2012dk,Chen:2014tpa}.
In review, this is not surprising as it has already been understood much earlier that 
outer automorphisms also describe CP transformations of gauge \cite{Grimus:1995zi} and space-time symmetries \cite{Buchbinder:2000cq}.

A crucial difference between discrete and continuous groups is that many discrete groups are inconsistent with outer automorphisms 
which correspond to physical CP transformations in generic models \cite{Chen:2014tpa}. For such groups it has been
found that CP violating phases are calculable and take quantized (``geometrical'') values.
This has been observed in models of explicit \cite{Chen:2014tpa,Branco:2015hea,
Varzielas:2015fxa} and spontaneous CP violation \cite{Branco:1983tn, Chen:2014tpa} alike. 
Moreover, very recently a model has been found in which geometrical CP violation is predicted in the dynamical breaking of 
a continuous gauge symmetry to a discrete subgroup, while an unbroken outer automorphism enforces $\theta=0$ 
for the topological vacuum angle of the gauge group \cite{Ratz:2016scn}.

The aim of this talk is to give a brief overview of the topic of predictive CP violation and outer automorphisms.
For this, the concept of an outer automorphism (a ``symmetry of a symmetry'') is discussed based on simple examples. 
Then it is explained how the well established CP transformation of the SM can be understood 
as a simultaneous outer automorphism transformation of gauge and space-time symmetries.
Extrapolating from the SM, a general definition of a physical CP transformation is given.
Based on this we discuss the existence of model independent CP transformations for the most interesting continuous groups,
as well as the absence of such transformations for a certain class of discrete groups.
Finally, we illustrate a toy model of spontaneous geometrical CP violation 
in which the outer automorphism is unbroken and enforces $\theta=0$.
For the ease of the presentation we will skip many mathematical details.

\section{What is an outer automorphism?}

Outer automorphisms are transformations that map a group to itself,
while not being part of the group themselves. 
Unfortunately, outer automorphisms are rarely treated in the standard introductory literature on group theory.
Some details for the case of Lie groups can be found in \cite{Fuchs:1997jv} while a pedagogical introduction for finite groups
has been given in \cite{Trautner:2016ezn}.
Here, we will only discuss a very simple example of an outer automorphism, which suffices to understand the 
basic idea.

Consider the symmetry group \Z3 which is generated by a single element $\mathsf{a}$ and the relation $\mathsf{a}^3=\mathsf{id}$.
A complete list of group elements is $\left\{ \elm{id},\elm{a},\elm{a}^2 \right\}$. \emph{Inner} automorphisms are conjugation maps
$\elm{g}\mapsto\elm{g}'=\elm{h}\elm{g}\elm{h}^{-1}$ (with $\elm{g},\elm{g}',\elm{h}\in G$) and they partition the group $G$ into conjugacy classes.
Skipping over mathematical rigor, \emph{outer} automorphisms (\Outs) can be thought of as conjugation maps $\elm{g}\mapsto\elm{g'}=\elm{u}\elm{g}\elm{u}^{-1}$ with elements $\elm{u}$ that are \emph{not} part of the group.
For the case of \Z3 there is a single \Out which can be generated by the transformation
\begin{equation}\label{eq:Z3Out}
 \mathrm{Out}(\Z3)~:~\quad\elm{a}~\mapsto~u(\elm{a})~=~\elm{a}^2\;.
\end{equation}
We note that this map induces a permutation of group elements within the list $\left\{ \elm{id},\elm{a},\elm{a}^2 \right\}$.
In general, \Outs act as permutation of group elements and conjugacy classes while leaving the unsorted list of group elements invariant. 
Also, \Outs do not change the structure (i.e.\ the multiplication algebra) of the group.  
This justifies to call outer automorphisms the non-trivial ``symmetries of a symmetry''.

In a concrete sense, \Outs act as a permutation of representations of a group, $\Out\,:\,\rep[_i]{r}\mapsto\rep[_j]{r}$ where $i,j$ label the irreducible representations (irreps) of the group.
In order to be consistent with the group structure, the corresponding representation matrix $U$ of an \Out
has to fulfill \cite{Fallbacher:2015rea, Trautner:2016ezn} 
\begin{equation}\label{eq:ConsistencyCondition}
 U\,\rho_{\rep{r'}}(\mathsf{g})\,U^{-1}~=~\rho_{\rep{r}}(u(\mathsf{g}))\;,\qquad\forall \mathsf{g}\in G\;,
\end{equation}
where $\rho_{\rep{r}}(\mathsf{g})$ denotes the matrix representation of $g$ in the representation $\rep{r}$,
and $u(\mathsf{g})$ is the abstract \Out transformation on the group elements which maps $\rep{r}\mapsto\rep{r'}$.
In particular, this can include the special case $\rep{r'}=\rep{r}^*$ \cite{Holthausen:2012dk}. 
After having made a basis choice for \rep{r} and \rep{r'}, $U$ is fixed by this relation up to a central element of the group and a global phase.

For the case of \Z3, the simultaneous action of $\Out(\Z3)\cong\Z2$ on group elements and representations is depicted in \Tabref{tab:Z3}.
The transformation \eqref{eq:Z3Out} turns out to be a complex conjugation \Out in the sense that $\rep[_i]{r}\mapsto\rep[_j]{r}=\rep[_i]{r}^*$ for all irreps of \Z3.  

\begin{table}
\caption{Character table of \Z3. The arrows illustrate the action of the unique \Z2 outer automorphism on the conjugacy classes and representations.
We use the definition $\omega:=\mathrm{e}^{2\pi \I/3}$.}
\label{tab:Z3}
\begin{center}
 \begin{tabular}{llll}
 \br
 $\Z3$ & $\mathsf{id}$ & $\mathsf{a}^{\tikz\node [coordinate] (h1) {};}$ & ${}^{\tikz\node [coordinate] (h2) {};}${$\mathsf{a}^2$} \\
 \mr 
  {} $\rep{1}$ &  1 & 1 & 1 \\
  \tikz\node [coordinate] (n1) {}; $\rep{1'}$ & 1 & $\omega$ & $\omega^2$ \\
  \tikz\node [coordinate] (n2) {}; $\rep{1''}$ &  1 & $\omega^2$ & $\omega$ \\
 \br
\end{tabular}
\end{center}
\begin{tikzpicture}[overlay]
         \draw [<->,red,thick] (n1) to [bend right=45] (n2);
         \draw [<->,red,thick] (h1) to [bend left=15] (h2);
 \end{tikzpicture}
\end{table}%

\subsection{Outer automorphisms of groups}

For any group, the possible \Outs are dictated by the structure of the group itself. 
For the specific cases of simple Lie groups and for finite groups some general statements can be made.
The \Outs of simple Lie groups are given by the symmetries of the Dynkin diagram of the corresponding Lie algebra.
In the case of finite groups, \Outs are symmetries of the character table under permutation of rows and columns (cf.\ \Tabref{tab:Z3}).
However, this relation is not $1:1$, meaning that there can be non-trivial \Outs which leave the character table invariant.
Vice versa, there can also be symmetries of the character table which do not correspond to an \Out. 
Especially for discrete groups, the number of distinct \Outs can easily surpass the order of the group.
A useful tool in finding \Outs of small finite groups is 
\texttt{GAP} \cite{GAP4}. Some examples for groups and their \Outs are given in \Tabref{tab:OutExamples}.

\begin{table}
\caption{Outer automorphisms of all simple Lie groups as well as for some randomly chosen examples of finite discrete groups. The action on representations denoted by $\rep[_i]{r}\to\rep[_j]{r}$ denotes 
some unspecified permutation of representations other than a complex conjugation outer automorphism.}
\label{tab:OutExamples}
\begin{center}
\begin{tabular}{lllll}
 \br
 Simple Lie Group & Algebra & Out & Action on reps   \\
 \mr  
    $\SU{N}$  & $A_{n>1}$ & $\Z2$ & $\rep{r}\,~\to~\rep{r}^*$  \\
    $\SO{8}$  & $D_{n=4}$ & $\mathrm{S}_3$ & $\rep{r}_i~\!\to~\rep{r}_j$ \\
    $\SO{2N}$ & $D_{n>4}$ & $\Z2$ & $\rep{r}\,~\to~\rep{r}^*$ \\
    $E_6$     & $E_6$     &  $\Z2$ & $\rep{r}\,~\to~\rep{r}^*$ \\ 
    all others    &  &   $\cancel{\phantom{1}}$ & $\cancel{\phantom{1}}$ \\
 \br
\end{tabular}\hspace{1cm}
\begin{tabular}{llll}
 \br
 Group & Out & Action on reps  \\
 \mr  
 $\Z{3}$ & $\Z2$ & $\rep{r}\,~\to~\rep{r}^*$  \\
 ${\rm A}_{n\neq6}$ & $\Z2$ & $\rep{r}\,~\to~\rep{r}^*$  \\
 ${\rm S}_{n\neq6}$ & $\cancel{\phantom{1}}$ & $\cancel{\phantom{1}}$  \\
 $\T7$ & $\Z2$ & $\rep{r}_i~\!\to~\rep{r}_j$  \\
 $\Delta(27)$ & ${\rm GL}(2,3)$ & $\rep{r}_i~\!\to~\rep{r}_j$ \\
 $\Delta(54)$ & ${\rm S}_{4}$ & $\rep{r}_i~\!\to~\rep{r}_j$ \\
 \br
\end{tabular}%
\end{center}
\end{table}%

\subsection{Physics of outer automorphisms}

A generic physical model (possibly a QFT) can be specified by stating the underlying symmetry group as well as the field content in terms of representations of the group.
The symmetry group may have a number of \Outs, and a central question is how these are related to the physics described by the model.

As described above, the general action of an \Out on the irreps of a group is a permutation map $\rep[_i]{r}\mapsto\rep[_j]{r}$.
For a model which includes an irrep $\rep[_i]{r}$ there are, in general, two logical possibilities for the action of an \Out:
\begin{itemize}
 \item[(i)] Either, both $\rep[_i]{r}$ and $\rep[_j]{r}$ are included in the model,
 \item[(ii)] or, $\rep[_j]{r}$ is not part of the model.
\end{itemize}
Let us briefly discuss these two cases, and give two well-known examples.

In the first case (i), the \Out transformation would, in principle, be a possible symmetry transformation of the model in the sense that it
permutes already present (symmetry invariant) operators.
However, as it is not a symmetry of the model \textit{by assumption}, it will generally be broken by the values of some couplings. 
Turning this around it is clear that the \Out will join the symmetries of the model \textit{iff} the couplings are tuned to very specific values.
Altogether, the \Out behaves very much like an explicitly (softly) broken symmetry transformation. 
Regions in the parameter space with enhanced symmetry, thus, can be identified by \Outs. The existence of such regions has interesting implications in general. 
For example, parameter values with enhanced symmetry will be boundaries of the RGE flow.

In the second case (ii), the \Out is in no way a possible symmetry of the model as the target representation $\rep[_j]{r}$ 
is not part of the model. Performing the \Out mapping nevertheless, would map the model to a physically distinct model. 
We see that the \Out is broken by the representation content of the model.
Transformations of this kind cannot be conserved by tuning of parameters 
and they are called explicitly and \emph{maximally} broken.

Let us give two examples for the possible action of \Outs based on the SM.
\begin{itemize}
 \item[(i)] CP transformation in the SM. Anticipating the results of the next section, 
 the CP transformation of the SM maps $\rep{r}\mapsto\rep{r}^*$ for all representations (details will be given below). 
 For example, it maps
 \begin{equation}
  Q_\mathrm{L}~\equiv~\left(\rep{3},\rep{2}\right)^{\rm{L}}_{1/6}~\mapsto~\left(\bar{\rep{3}},\bar{\rep{2}}\right)^{\rm{R}}_{-1/6}~\sim~Q_\mathrm{L}^*\;. 
 \end{equation}
 Performed on the action, this transformation is equivalent to a mapping in the parameter space $V_{\rm CKM}\mapsto \left(V_{\rm CKM}\right)^*$.
 Is this a symmetry of the model? Only experimental input can decide this question. Experimentally it is well established that 
 $\delta_{\rm CKM}\neq\left\{0,\pi\right\}$, for what reason this \Out, which corresponds to a physical CP transformation, is explicitly broken by the values of couplings. 
\item[(ii)] P transformation in the SM. Pictorially, parity maps $\rep[_{\rm L}]{r}\mapsto\rep[_{\rm R}]{r}$ where $\rm{L}$ and $\rm{R}$ 
denote chiral representations of the Lorentz-group. In the SM, for example 
  \begin{equation}
   Q_{\rm{L}}~\equiv~\left(\rep{3},\rep{2}\right)^{\rm{L}}_{1/6}~\mapsto~\left(\rep{3},\rep{2}\right)^{\rm{R}}_{1/6}~\equiv~Q_{\rm{R}}\;.
  \end{equation}
  However, $\rm{Q}_R$ is not part of the model to begin with, meaning that this transformation would map the SM to some other model.
  We see that this transformation is broken explicitly and \emph{maximally}, i.e.\ not by the values of couplings but by the representation content 
  of the model.
\end{itemize}
Even though these examples may appear trivial, they nicely illustrate the possible action of \Outs in general, also for other models
and also for transformations other than the well-known P or CP.
For a first application of this to models with \Outs other than CP see \cite{Fallbacher:2015rea, Trautner:2016ezn}.
As the concept of symmetries of symmetries is as general as the concept of symmetry, 
we expect that many more applications of such transformations will come to light in the future. 

With this formulation at hand, it is very easy to argue that CP cannot be broken maximally, i.e.\ by leaving out or adding representations.
For a real valued action, the Lagrangian must always involve both, $\rep{r}$ and $\rep{r}^*$ in mutually hermitean conjugate operators. 
Therefore, a simultaneous mapping $\rep{r}\mapsto\rep{r}^*$ of all representations cannot be broken maximally. 

\section{CP as a special (outer) automorphism}

Let us show that the usual CP transformation of the SM is a simultaneous complex conjugation \Out of all present symmetries.

The most general possible CP transformation for (SM) gauge and one generation of (chiral) fermion fields is given by (cf.\ e.g.\ \cite{Grimus:1995zi}) 
\begin{equation}\label{eq:MostGeneralCP}
\begin{split}
W^a_\mu(x)~&\mapsto~R^{ab}\,\parityP_\mu^{\;\nu}\,W^b_\nu(\parityP\,x)\;, \\
\Psi_{\alpha}^i(x)~&\mapsto~\eta_\mathsf{CP}\,U^{ij}\,\mathcal{C}_{\alpha\beta}\,{\Psi^*}_{\beta}^j(\parityP\,x)\;.
\end{split}
\end{equation}%
Here $W^a_\mu$ and $\Psi_{\alpha}^i$ denote the gauge and fermion fields, $\parityP=\mathrm{diag}(1,-1,-1,-1)$ is the usual spatial reflection, and $\eta_\mathsf{CP}$
denotes an arbitrary phase. Greek lowercase letters run within the corresponding Lorentz group representation, 
while lowercase Roman letters run within the respective gauge group representation and we implicitly sum over repeated indices. 
$\mathcal{C}$ and $U$ here are general unitary matrices, allowing the fermions to rotate in the Dirac representation space of the Lorentz group 
and in the representation space of the gauge group, respectively. $R$ is orthogonal and can be chosen to be real  
due to the reality of the gauge fields.

The CP transformation \eqref{eq:MostGeneralCP} is a conserved symmetry of the gauge-kinetic action \emph{iff}
\begin{align}\nonumber
(\mathrm{i})~&:~&R_{aa'}\,R_{bb'}\,f_{a'b'c}~&=~f_{abc'}\,R_{c'c}\;,& \\ \label{eq:CPasOut2}
(\mathrm{ii})~&:~&U\,(-\elm{T}^\mathrm{T}_a)\,U^{-1}~&=~R_{ab}\,\elm{T}_b\;,& \\ \nonumber\label{eq:CPasOut3}
(\mathrm{iii})~&:~&\mathcal{C}\,(-{\gamma^\mu}^\mathrm{T})\,\mathcal{C}^{-1}~&=~{\gamma^\mu}\;.&
\end{align}
Here, $\elm{T}_a$ and $f_{abc}$ are the generators and structure constants of the gauge group fulfilling the usual relation 
$\left[\elm{T}_a,\elm{T}_b\right]=\I f_{abc}\,\elm{T}_c$, while $\gamma^\mu$ denote the usual Dirac $\gamma$-matrices which form the generators $S^{\mu\nu}=\I/4\times\left[\gamma^\mu,\gamma^\nu\right]$ of the Dirac spinor representation.

The equations \eqref{eq:CPasOut2} are special cases of the general consistency condition \eqref{eq:ConsistencyCondition} for the gauge
and space-time symmetries.  
This shows that the most general physical CP transformation is a special automorphism of
both, the gauge and space-time symmetries of a model. Namely, the physical CP transformation in the SM
corresponds to an automorphism which simultaneously maps all present (gauge and space-time) symmetry representations
to their own complex conjugate representations, $\rep[_i]{r}\mapsto\rep[_i]{r}^*~\forall~i$.
Such an automorphism must be outer for a group that has complex representations. 
To prevent confusion we emphasize that neither C nor P alone are complex conjugation automorphisms of the Lorentz group. 
Only the combined CP transformation is a complex conjugation \Out of the Lorentz group (cf.\ e.g.\ \cite{Buchbinder:2000cq}).

If the field $\Psi$ would, in addition, also transform under some additional global symmetry, it is imperative that 
also the corresponding representation of the global symmetry is mapped onto its own complex conjugate representation.
That is, also for all additional symmetry groups beyond the SM the CP transformation should map representations to their own complex conjugate representations.\footnote{%
Strictly speaking this is true only for fields which are charged under the SM gauge group or which couple to SM neutral operators that are charged under the new 
symmetry, implying that completely neutral fields or completely decoupled sectors could behave differently.}
Extrapolating from the SM, we can use these insights in order to formulate a group theoretical definition of a physical CP transformation.
In words:

\begin{center}
\textit{A physical CP transformation is an (outer) automorphism transformation which simultaneously maps all present representations of all symmetries (global, local, space-time) to their respective complex conjugate representations.}
\end{center}

This definition generalizes previous definitions of CP given for simple Lie groups \cite{Grimus:1995zi} or the Poincar\'e group \cite{Buchbinder:2000cq}.
Of course, it is also consistent with the standard textbook treatments of CP.

We note that a physical CP transformation does not have to be unique (i.e.\ there can be multiple, equally valid CP transformations) nor is it guaranteed to exist at all for a given symmetry.

As an important remark, we note that it is, of course, 
always possible to impose \textit{any} complex conjugation (i.e.\ physical CP) transformation on a given model -- even if there exists no corresponding complex conjugation automorphism of the symmetry group $G$ of the model.
However, this will generally enhance the linear symmetry of the model, possibly up to a decoupling of the model (as it happens for our example in \Secref{sec:SU3T7}). 
It would then be unreasonable to speak of a $G$-symmetric setting for what reason this case is not considered here.

Let us also remark that there are, in general, \Outs which have nothing to do with physical CP whatsoever.
In general, it is model dependent whether or not a given automorphism constitutes a physical CP transformation.
For example, an automorphism which corresponds to a physical CP transformation for one model may not be a physical CP transformation
once additional representations are added to the model. This will be the case for discrete groups of type I below.
In contrast, if a symmetry allows for a complex conjugation automorphism, this always corresponds to a model independent physical CP transformation. 

Finally, we stress that our above definition does not require the CP transformation to be of order two, i.e.\ it equally allows for CP transformations of higher order.

\section{Do CP transformations exist for all models/symmetry groups?}
Following the previous section we understand model independent physical CP transformations as complex conjugation \Outs of a group, and 
the general answer to the question in the subheading is no.

\subsection{CP transformations for Lie groups}

For the particular case of (semi-)simple Lie groups it has been shown that complex conjugation \Outs always exist \cite{Grimus:1995zi}. 
This does, of course, not imply that CP is automatically conserved in a model based on such a group.
In contrast, models which allow for CP transformations are generally not predictive with respect to the CP violating phase,
meaning that it would need an experiment to find out whether or not CP is violated explicitly.
There are, however, situations in which all potentially CP violating parameters are unphysical 
and CP is automatically conserved (cf.\ \cite{Haber:2012np} for a very detailed treatment of when this is the case).

For other groups, it has to be checked on a case-by-case basis whether they allow for a 
complex conjugation automorphism transformation or not. The physically interesting case of the Poincar\'e 
or Lorentz group has been investigated in \cite{Buchbinder:2000cq} and it has been found that there exists a complex conjugation 
\Out which serves as a model independent physical CP transformation.

\subsection{CP and discrete groups}

A prominent set of groups for which complex conjugation \Outs do not, in general, exist are (finite) discrete groups \cite{Chen:2014tpa}.
A simple indicator to decide whether or not a complex conjugation \Out exists for a given discrete group has been presented in \cite{Chen:2014tpa} (see also \cite{Bickerstaff:1985jc}).
In particular, all (finite) discrete groups can be classified into two categories depending on whether or not a complex conjugation \Out exists.
We will briefly discuss the categories in the following. Examples for each category of finite groups are shown in \Tabref{tab:groupTypes}.

\begin{table}
\caption{Some examples for discrete groups of type I, II~A, and II~B with their common names and SmallGroups library ID of GAP, see \cite{Chen:2014tpa} for details.}
\label{tab:groupTypes}
\begin{center}
\subtable[Examples for groups of type I.
 \label{tab:typeI}]
 {\begin{tabular}{lllll}
 \br
 $\boldsymbol{G}$ & $\Z5 \rtimes \Z4$ & $\mathrm{T}_7$  & $\Delta(27)$ & $\Z9 \rtimes \Z3$  \\
 \mr
 SG id & $(20,3)$ & $(21,1)$ & $(27,3)$ & $(27,4)$ \\
 \br
 \end{tabular}} \\
\subtable[Examples for groups of type II~A.\label{tab:typeIIA}]
{\begin{tabular}{llllll}
 \br
  $\boldsymbol{G}$ & $S_3$ & $A_4$ & $\mathrm{T}'$ & $S_4$ & $A_5$\\
  \mr
  SG id & $(6,1)$ & $(12,3)$ & $(24,3)$ & $(24,12)$ & $(60,5)$\\
  \br
  \end{tabular}}\\
\subtable[Examples for groups of type II~B.\label{tab:typeIIB}]
  {\begin{tabular}{lll}
  \br
  $\boldsymbol{G}$ & $\Sigma(72)$ & $\left((\Z3 \times \Z3) \rtimes \Z4\right) \rtimes \Z4$\\
   \mr
   SG id & $(72,41)$ & $(144,120)$ \\
   \br
  \end{tabular}}
\end{center}  
\end{table}%

Groups of the so-called ``type I'' do not allow for a complex conjugation \Out. 
Depending on the representation content of a specific model based on a type I group, it can happen that the physical CP transformation $\rep[_i]{r}\mapsto\rep[_i]{r}^*~\forall~i$ is inconsistent
with the symmetry of the model. If this is the case, CP is violated \emph{as a consequence} of requiring the type I symmetry.
Explicitly it has been found that the necessarily complex
Clebsch--Gordan coefficients of type I groups then enter CP violating observables such as asymmetries in
oscillations and decays \cite{Chen:2014tpa}.
These phases can be calculated and they take discrete ``geometrical'' values (fractions of $2\pi$).
Models based on type I groups, hence, can be truly predictive with respect to CP violation.
This way of CP violation has been termed explicit geometrical CP violation \cite{Chen:2009gf,Branco:2015hea,
Varzielas:2015fxa}.

We stress that not all models based on type I groups automatically violate CP. 
In non-generic settings CP can be conserved even though the model features a type I symmetry.
This can happen only if the group has an automorphism that maps $\rep[_j]{r}\mapsto\rep[_j]{r}^*$ 
for a subset of representations $j$ (including a faithful representation), and if the model contains only representations out of this subset.

While CP then can be conserved, it could also be violated spontaneously. 
In this case, the discrete phases originating from the complex Clebsch-Gordan coefficients 
of the type I group can become relative physical phases of VEVs \cite{Fallbacher:2015rea}. 
This is known as spontaneous geometrical T violation \cite{Branco:1983tn}.

The second category of finite groups are groups of ``type II''. 
These groups allow for an \Out that simultaneously maps all irreps to their own complex conjugate irreps. 
This allows for a model independent definition of a physical CP transformation.
CP violation in type II groups then works completely analogous to the case of simple Lie groups. 
Namely, it is always possible to perform a CP transformation in consistency with the symmetry.
The CP transformation could be violated by the value of certain couplings, but CPV cannot be predicted.

A special subset of type II groups are the groups of the so-called type II~B. For them, a complex conjugation \Out exists, 
but it generates a CP transformation of order larger than two. 
In this case, CP can be defined but conserving it typically requires an additional global symmetry \cite{Chen:2014tpa}.

\subsection{Higher order CP transformations}
As an aside we note that the presence of complex couplings is in general \emph{not} sufficient to conclude
that CP is violated. More specifically, it has been found that there are models with higher-order CP transformations
in which couplings are necessarily complex even though CP is manifestly conserved \cite{Ivanov:2015mwl}.
Exactly the same situation arises for groups of the type II~B, which also have only higher-order CP transformations and 
necessarily complex Clebsch-Gordan coefficients entering couplings \cite{Chen:2014tpa}. 
A generic consequence of conserved higher-order CP transformations is the presence of additional linear symmetries \cite{Grimus:1987kn}
as well as the presence of exotic CP ``half--odd'' states with potentially very 
interesting phenomenological consequences \cite{Aranda:2016qmp}.

\section{CP violation in the breaking of continuous to discrete}
\label{sec:SU3T7}

An interesting observation is that type I groups can arise as subgroups of
simple Lie groups. 
While simple Lie groups allow for a complex conjugation \Out
which serves to define a model independent physical CP transformation, a type I subgroup does not have 
a complex conjugation \Out and, hence, also no possible physical CP transformation in a generic setting.
As argued above, this implies that CP is violated in generic settings based on a type I group.
An immediate question, therefore, is what 
happens to the CP transformation if a simple Lie group is broken to a type I subgroup.

This has been investigated in a concrete model \cite{Ratz:2016scn} based on a gauged \SU3 symmetry which is dynamically
broken to a \T7 subgroup by the vacuum expectation value (VEV) of a complex scalar $\phi$ in the $\rep{15}$-plet representation of \SU3 (Dynkin indices $\left\{2,1\right\}$).
The model is given by
\begin{equation}
\mathscr{L}~=~\left(D_\mu\,\phi\right)^\dagger\left(D^\mu\,\phi\right)
-\frac{1}{4}\,G^a_{\mu\nu}\,G^{\mu\nu,a}-\mathscr{V}(\phi)\;,
\end{equation}
where $D_\mu=\partial_\mu-\I g\,A_\mu$ is the gauge covariant
derivative and $G_{\mu\nu}^a$ the field strength tensor. 
The scalar potential is given by
\begin{equation}
 \mathscr{V}(\phi)~=~-\mu^2\,\phi^\dagger\phi+\sum^5_{i=1}\lambda_i\,\mathcal{I}^{(4)}_i(\phi)\;,
\end{equation}
with $\mu^2>0$, and $\mathcal{I}^{(4)}_i$ denote the five
possible quartic \SU3 invariants in the self-contraction of the $\rep{15}$-plet. These invariants are real and hence $\lambda_i\in\mathbbm{R}$ without loss of generality.
For some choice of parameters, this
potential has a global minimum $\langle\phi\rangle$
that spontaneously breaks $\SU3\rightarrow\T7$ \cite{Luhn:2011ip, Merle:2011vy}.
Crucially, the VEV does \emph{not} break the \Out transformation of \SU3. 
Thus, if the \Out was imposed to be an exact symmetry initially, it continues to 
be an exact symmetry also in the broken phase. 

Requiring CP conservation at the level of \SU3 amounts to require the complex conjugation automorphism to be conserved. 
The actual symmetry breaking chain in the model is then given by
\begin{equation}
  \SU3\rtimes\Z2~\xrightarrow{\langle\phi\rangle}~\T7\rtimes\Z2\;.
\end{equation}
However, at the level of \T7, the \Z2 \Out
transformation does not correspond to a physical CP transformation.
This can be understood because some of the physical (\T7) states in the broken phase 
are complex linear combinations of \SU3 states, in such a way that 
the \Out does not map them to their own complex conjugate.

At the level of \T7, physical CP is thus violated by calculable phases, irrespective of the presence 
of an unbroken \Out \cite{Ratz:2016scn}. This has been explicitly cross-checked by computing a non-vanishing CP-odd basis invariant in the broken phase of the model \cite{Ratz:2016scn}. 

Note that there is also no other possible CP transformation at the level of \T7.
This can easily be concluded from the fact that \T7 is of type I, and that the branching of the \SU3 fields to \T7 is such that all \T7 irreps are populated.
If one would plainly enforce a CP transformation on the model by simply requiring the transformation $\rep{r}_{\T7}\mapsto\rep{r}_{\T7}^*$
for all \T7 representations to be a symmetry, all couplings vanish and one ends up with a free theory. 

The conserved \Out acts on the \SU3 gauge fields like a CP transformation,
i.e.\ it prohibits a possible topological term
\begin{equation}\label{eq:Ltheta}
\mathscr{L}_\theta~=~\theta\,\frac{g^2}{32\pi^2}\,G^a_{\mu\nu}\,\widetilde{G}^{\mu\nu,a}\;.
\end{equation}
The unbroken \Out continues to prohibit the $\theta$-term in the broken phase, where the \Out, however, cannot be any longer interpreted as a physical CP transformation
of the \T7 states.
This provides a symmetry reason for $\theta=0$ also in the broken phase, despite the fact that CP is violated by other \T7 invariant couplings \cite{Ratz:2016scn}.

Whether or not a similar construction can serve as a solution to the strong CP problem of the SM remains to be seen. 
In any case it is interesting to see
that one and the same outer automorphism can have different physical interpretations depending on the 
linearly realized symmetries of the ground state of a theory.

\section{Conclusion}

It has been shown how outer automorphisms can be understood as non-trivial symmetries of a symmetry.
For physics purpose these transformations are best thought of as (certain) permutations 
among the irreps of a group.
We have then shown that CP transformations in the SM are 
complex conjugation outer automorphisms of gauge and space-time symmetries.
Extrapolating from the SM, we have given a general definition of a physical CP transformation
as a simultaneous mapping of all symmetry representations to their own respective complex conjugate representations.
It has been pointed out that physical CP transformations are clashing with certain discrete symmetries,
in which case CP violating phases with quantized values can be predicted.
Finally, we have presented a toy model in which a continuous gauge symmetry is dynamically broken to a discrete subgroup that 
does not allow for a physical CP transformation.
In this case geometrical CP violation arises while an unbroken outer automorphism transformation warrants $\theta=0$.

\section*{Acknowledgments}
This work has been supported by the German Science Foundation (DFG) within the SFB-Transregio TR33 "The Dark Universe".

\bibliographystyle{iopart-num}
\section*{References}
\bibliography{Orbifold}

\end{document}